\documentclass[%
 reprint,
 amsmath,amssymb,
 aps,
]{revtex4-2}

\usepackage{graphicx}
\usepackage{dcolumn}
\usepackage{bm}
\usepackage{hyperref}
\usepackage{lipsum} 
\usepackage{xcolor}
\usepackage{braket}
\usepackage{booktabs}
\usepackage{upgreek}
\usepackage{siunitx}

\DeclareMathOperator{\Var}{Var}
\newcommand\Trot{\ensuremath{T_{\mathrm{rot}}}}
\newcommand\navg{\ensuremath{\langle{}n\rangle}}

\begin{document}

\preprint{APS/123-QED}

\title{Size characterization of neutral rare-gas clusters based on time-resolved polarization anisotropy measurements}

\author{Arne Morlok}
\thanks{These two authors contributed equally}
\affiliation{Institute of Physics, Universitiy of Freiburg}
\author{Grzegorz Kowzan}
\thanks{These two authors contributed equally}
\affiliation{Institute of Physics, Nicolaus Copernicus University in Toru\'n}
\author{Yilin Li}
\affiliation{Institute of Physics, Universitiy of Freiburg}
\author{Ulrich Bangert}
\affiliation{Institute of Physics, Universitiy of Freiburg}
\author{Frank Stienkemeier}
\affiliation{Institute of Physics, Universitiy of Freiburg}
\author{Lukas Bruder}
\email{lukas.bruder@physik.uni-freiburg.de}
\affiliation{Institute of Physics, Universitiy of Freiburg}
\date{\today}

\begin{abstract}
  The size determination of neutral clusters is experimentally challenging.
  In particular, weakly-bound rare-gas clusters tend to fragment upon ionization, resulting in systematic errors in cluster size studies.
  In contrast, characterization of the temporal polarization anisotropy dephasing provides a soft detection scheme for cluster size estimation, which avoids fragmentation of the clusters.
  Here, we present a systematic experimental study of argon and neon clusters in the size range of 50 to 10.000 atoms using this technique.
  In order to extract the mean cluster sizes from the data, we present an efficient analytical model of the polarization anisotropy dephasing of an ensemble of doped clusters.
  The approach shows remarkable sensitivity to small changes in the mean cluster size of just a few tens of atoms and allows us to refine the widely used Hagena scaling law for the estimation of rare-gas cluster sizes.
\end{abstract}
\maketitle

\section{\label{sec:Intro}Introduction}
Van der Waals-bound rare-gas clusters are a unique and widely employed sample environment to study fundamental properties of isolated atoms and molecules\cite{Mestdagh1997,Toennies2004,Choi2006,Kuepper2007,Stienkemeier2006,Briant2022}.
In a variety of experiments the control and precise knowledge of the mean cluster size is critical \cite{Lallement1992,Lallement1993,Muller2015,Izadnia2017,Bohlen2022,Briant2022}.
One example is cluster-isolated chemical reactions (CICR)\,\cite{Mestdagh1997,Farnik2021,Briant2022}.
Here, rare-gas clusters serve as inert and cold nanoreactor for the study of isolated chemical reactions \cite{Lallement1992,Lallement1993,Briant2022}.
Furthermore, cluster size is key to controlling and tuning collective inter-molecular processes of adsorbates on the cluster surface such as singlet fission or superradiance\cite{Muller2015,Izadnia2017,Bohlen2022}.

Despite its importance, experimentally measuring the cluster size remains challenging, in particular for neutral rare-gas and other neutral cluster species \cite{Lee2020,Pandey2021,Yao2023}. Seminal studies to determine scaling laws for the mean cluster size of rare-gas clusters prepared by supersonic expansion have been conducted by Hagena and coworkers employing mass spectrometry based on electron impact ionization \cite{Hagena1972,Hagena1974,Hagena1981,Hagena1992}. Their work showed that the expansion conditions can be combined into a single semi-empirical parameter $\Gamma$ with which the mean cluster size can be predicted. $\Gamma$ is given by
\begin{equation}
  \label{eq:gamma}
  \Gamma = \dfrac{p_0}{k_B} d^qT_0^{0.25q - 2.5},
\end{equation}
where $p_0$ is the expansion pressure, $T_0$ is the expansion temperature, $d$ is the effective nozzle opening diameter and $q$ is an empirical parameter for which Ref.~\cite{Hagena1987} suggests a value of $q = 0.85$ for argon clusters. In order to account for the interaction potential of different gases, $\Gamma$ is normalized with a gas-characteristic constant $\Gamma_{\mathrm{ch}}$. This leads to a dimensionless quantity $\Gamma^* = \Gamma / \Gamma_\mathrm{ch}$ with which the mean cluster size $\navg$ can be determined according to \cite{Hagena1992}
\begin{align}
  \label{eq:Hagena}
  \navg = 33 \cdot \left(\dfrac{\Gamma^*}{1000}\right)^{2.35}.
\end{align}
It should be noted that these scaling relations are valid for continuous jet expansion of rare gases except for helium, which forms superfluid helium nanodroplets upon supersonic expansion \cite{Slenczka2022}. Because of their exceptional properties, helium nanodroplets are not considered in this study. Likewise, pulsed gas expansion is not considered here.

Although further research confirmed the general cluster size scaling behavior found by Hagena and coworkers \cite{Farges1986,Karnbach1993,Buck1996,Yao2023}, deviations have been reported when measuring the mean cluster size with different methods, including electron scattering and diffractive He atom scattering \cite{Farges1986,Buck1996}, Rayleigh scattering \cite{Lee2020} as well as mass spectrometry with laser-based ionization \cite{Yao2023}. Buck and Krohne, for example, suggested an adjustment to the Hagena model for small clusters \cite{Buck1996}. In case of $350 \le \Gamma^* \le 1800$, the mean cluster size is more accurately obtained with
\begin{equation}
  \label{eq:Buck}
  \navg = 38.4 \cdot \left( \dfrac{\Gamma^*}{1000}\right)^{1.64}.
\end{equation}
Eqs. \ref{eq:Hagena} and \ref{eq:Buck} are widely used for the size estimation of rare-gas clusters which we term Hagena scaling laws in the remainder of the paper.
The reason for the deviations between Eq.\,\ref{eq:Hagena} and Eq.\,\ref{eq:Buck} is most likely systematic errors in the determination of cluster size distributions.
In the Hagena experiments, this can be caused by cluster fragmentation upon electron impact ionization, which is common for rare-gas clusters \cite{Buck1986,Schutte2002}
Hence, soft measurement techniques where little energy is transferred to the clusters should be preferred for cluster-size determination. In this context, Yao and coworkers recently presented a facile soft ionization approach where rare-gas clusters are doped with aromatic molecules that are selectively photoionized\cite{Yao2023}. Fragmentation is prevented in this approach, as the cluster is neither directly excited nor ionized. Although their results generally reproduced the results of the Hagena model, they found an unexpected deviation regarding the scaling behavior of cluster size with temperature and pressure, highlighting the necessity for further work on improving the precision of the Hagena scaling laws.

Here, we present a novel ionization-free approach to measure the mean cluster size of neutral rare-gas cluster beams.
To this end, we surface-dope the clusters with molecular chromophores and characterize the cluster size distribution by measuring the transient alignment and ensemble dephasing of the molecular transition dipoles.
The approach is related to rotational coherence spectroscopy (RCS) introduced by Hochstrasser \textit{et al.}~\cite{Myers1986a} and by Zewail \textit{et al.}~\cite{Felker1987,Baskin1987,Baskin1994}.
A similar approach to ours has been recently demonstrated in a study of solvation dynamics in doped Ar clusters~\cite{Awali2021}.
We extend on this by providing an improved, physically justified theoretical model and a systematic study of cluster size distributions and cluster species.

\section{\label{sec:Methods}Methods}
\subsection{Experiment}
\begin{figure}
  \centering
  \includegraphics[width=\linewidth]{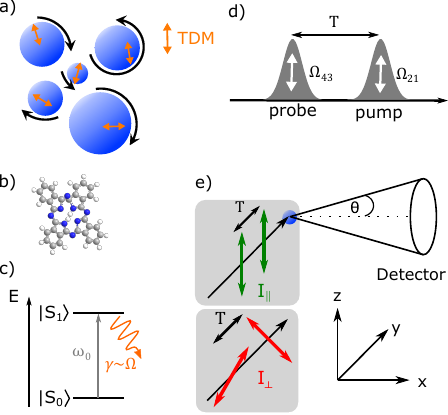}
  \caption{Experimental scheme. \textbf{a} Doped rare-gas clusters. Orange arrows indicate the direction of the dopant's transition dipole moment (TDM), black arrows the cluster rotation. \textbf{b} H$_2$Pc molecule. \textbf{c} Relevant electronic levels of H$_2$Pc and fluorescence yield modulated at $\Omega$ (see \textbf{d} and main text). \textbf{d} Pump-probe modulation scheme. Pump and probe pulses are amplitude-modulated at frequencies $\Omega_{21, 43}$, respectively. $T$ denotes the pump-probe delay. \textbf{e} Excitation geometry for excitation either with parallel ($I_\parallel$, green) or perpendicular ($I_\perp$, red) pump-probe pulse polarization, from which the polarization anisotropy parameter $r(T)$ is determined. $\theta^\prime$ refers to the opening angle of the detector.}
  \label{fig:experimental_scheme}
\end{figure}
\begin{table}
  \centering
  \caption{\label{tab:cluster_para} Expansion conditions indicating nozzle temperature $T_0$ and stagnation pressure $p_0$ (compare Eq.\,\ref{eq:gamma}. In case of neon, a nozzle with an opening diameter of 5\,$\upmu$m was used, and in the case of argon, a nozzle with an opening diameter of 10\,$\upmu$m. The mean cluster sizes $\langle n_\mathrm{Hag} \rangle$ are calculated according to the Hagena scaling laws.}
  \begin{tabular}{lccc}
    \toprule
    Species & $T_0$ (K) & $p_0$ (bar) & $\langle n_\mathrm{Hag} \rangle$ \\
    \midrule
    Ar      & 296       & 50          & 59                               \\
    Ar      & 296       & 70          & 136                              \\
    Ar      & 296       & 90          & 246                              \\
    Ar      & 232       & 70          & 499                              \\
    Ar      & 217       & 70          & 729                              \\
    Ar      & 207       & 75          & 1091                             \\
    Ar      & 190       & 70          & 1470                             \\
    Ar      & 191       & 80          & 1984                             \\
    Ne      & 85        & 50          & 67                               \\
    Ne      & 65        & 50          & 307                              \\
    Ne      & 58        & 60          & 870                              \\
    \bottomrule
  \end{tabular}
\end{table}
Supersonic beams of Ne and Ar clusters of varying size are prepared in a continuous gas expansion.
The vacuum apparatus has been described elsewhere \cite{Bruder2019}.
The expansion conditions are given in Tab.\,\ref{tab:cluster_para}.
The clusters are doped with free-base phthalocyanine (H$_2$Pc, see Fig.~\ref{fig:experimental_scheme}b) molecules of which the S$_1$$\leftarrow$S$_0$ transition is excited (Fig.~\ref{fig:experimental_scheme}c).
  The optical resonances are at $\lambda_\mathrm{Ar} = 674$\,nm on Ar clusters and $\lambda_\mathrm{Ne} = 665$\,nm on Ne clusters.
  Hence, dipole transitions in the visible domain can be probed which avoids the need for vacuum ultraviolet radiation to directly interact with the rare-gas species.

  We use a femtosecond pump-probe scheme (Fig.~\ref{fig:experimental_scheme}d,e) to excite a polarization in the ensemble and record the anisotropy dephasing of this polarization as a function of the pump-probe delay $T$.
  In two consecutive measurements, the fluorescence yield from the doped clusters is detected either for parallel polarization of the pump and probe pulses yielding the signal intensity $I_\parallel(T)$ and perpendicular polarization yielding $I_\perp(T)$.
  The polarization anisotropy parameter $r(T)$ is calculated according to
  \begin{align}
    \label{eq:ani_exp}
    r(T) = \dfrac{I_\parallel(T) - I_\perp(T)}{I_\parallel(T) + C I_\perp(T)}.
  \end{align}
  The parameter $C$ accounts for the normalization of the anisotropy signal with respect to the excitation and detection geometry, particularly the detector opening angle $\theta$. The value of $C$ is determined analogously to the procedure in Ref.~\cite{Brown1999}, yielding a value of $C = 2.23$ for an opening angle of $\theta = 49.5$\thinspace$^\circ$. Details about the calculations can be found in the Supp. Info.

  %

  The experiment requires discrimination between the weak non-linear pump-probe signal from the much stronger linear fluorescence induced upon interaction with either pump or probe pulse only.
  This becomes particularly important in light of the low target densities in doped cluster beams (typ. $\sim 10^8$\,cm$^{-3}$).
  The selective detection is achieved by rapid amplitude modulation of the pump and probe pulses combined with lock-in amplification.
  To this end, we employ a collinear pulse train of four laser pulses of which the delay between pulse 1 and 2 as well as 3 and 4 is kept at zero. Hence, the pulse train effectively consists of two pulses (pump and probe, Fig.\,\ref{fig:experimental_scheme}d).
  The carrier-envelope phase of each of the four pulses is modulated with acousto-optical modulators driven each at a different modulation frequency (details about the setup can be found in Ref.\,\cite{Bruder2019}).
  The interference between the temporally overlapping pulses 1,2 and 3,4 results in a rapid amplitude modulation controlled by the relative frequency $\Omega_{21}$ and $\Omega_{43}$ between the acousto-optical modulators.
  The lock-in amplifier is tuned to the sum and difference frequency signal at $\Omega = \Omega_{21}+\Omega_{43}=13$\,kHz or $\Omega = \Omega_{21}-\Omega_{43}=3$\,kHz.
  This ensures, that only the nonlinear pump-probe signal is amplified and linear background is removed from the signal.
  We note that amplitude modulation of the pump and probe pulses could be also achieved with other, simpler approaches. We exploited here an existing optical setup for coherent two-dimensional spectroscopy\,\cite{Bruder2019}, where this type of amplitude modulation was readily implemented.

  The quality of the polarization conditions were investigated by determining the ratio between the intensity at the prepared polarization axis and the intensity at an angle of 90\,$^\circ$ to it. For the $I_\parallel$ geometry a ratio of at least 100:1 could be obtained and for the $I_\perp$ geometry a ratio as high as 10:1 was obtained. The reduced contrast for the $I_\perp$ geometry is due to the beam splitters employed in the optical interferometers of our setup.
  We correct for the resulting polarization ellipticity in the data analysis, as described below.

  \subsection{\label{sec:theory}Theory}

  In order to link the experimentally deduced anisotropy parameter $r$ (Eq.\,\ref{eq:ani_exp}) to the cluster size, we derive a theoretical model for the polarization anisotropy of an ensemble of doped, rotating clusters.
  The optical excitation with a femtosecond pump pulse creates a superposition of rotational states in the upper electronic state and fixes their phase.
  Since the pulse is spectrally broad, it transfers the equilibrium rotational state distribution from the ground state without reshaping it.
  In the subsequent free evolution, rotational quantum beats oscillate in a broad range of frequencies and rapidly go out of phase.
  These rotational beat frequencies correspond to classical spin rates.
  In fact, because the distribution of beat frequencies is wholly determined by thermal occupation of energy levels in the ground state, it matches the classical equilibrium distribution of angular velocities of a rigid rotor.

  Our method exploits the dependence of the angular velocity spectrum on cluster size to characterize the size distributions.
  Note that RCS~\cite{Myers1986a, Felker1987,Baskin1987,Baskin1994} usually studied distinct molecular species instead of broad distributions of similarly-shaped clusters.
  In the former case, $r(T)$ contains sharp rotational revivals, from which one can deduce molecular structures and rotational constants.
  As we will see, in the latter case this approach is not available.
  Nevertheless, for known rotational temperature of the ensemble, the mean cluster size can be determined from the initial anisotropy dephasing.
  This is because the angular velocity spectrum narrows with increasing cluster size, which implies that larger clusters stay in phase longer after initial excitation.

  The polarization anisotropy parameter is defined as:
  \begin{equation}
    \label{eq:r-definition}
    r(T; \Trot, \navg)=\frac{2}{5}\langle P_2[\hat{\mu}(0)\cdot \hat{\mu}(T)]\rangle,
  \end{equation}
  which is the ensemble-averaged two-time correlation function of the transition dipole directions.
  It describes time evolution of cluster orientation independent of any vibrational or electronic dynamics.
  We are explicitly interested in the dependence of $r$ on time, $T$, rotational temperature, $\Trot$, and average cluster size, $\navg$, but we will omit the latter two variables for brevity.

  \begin{figure}[ht]
    \centering
    \includegraphics[width=0.8\linewidth]{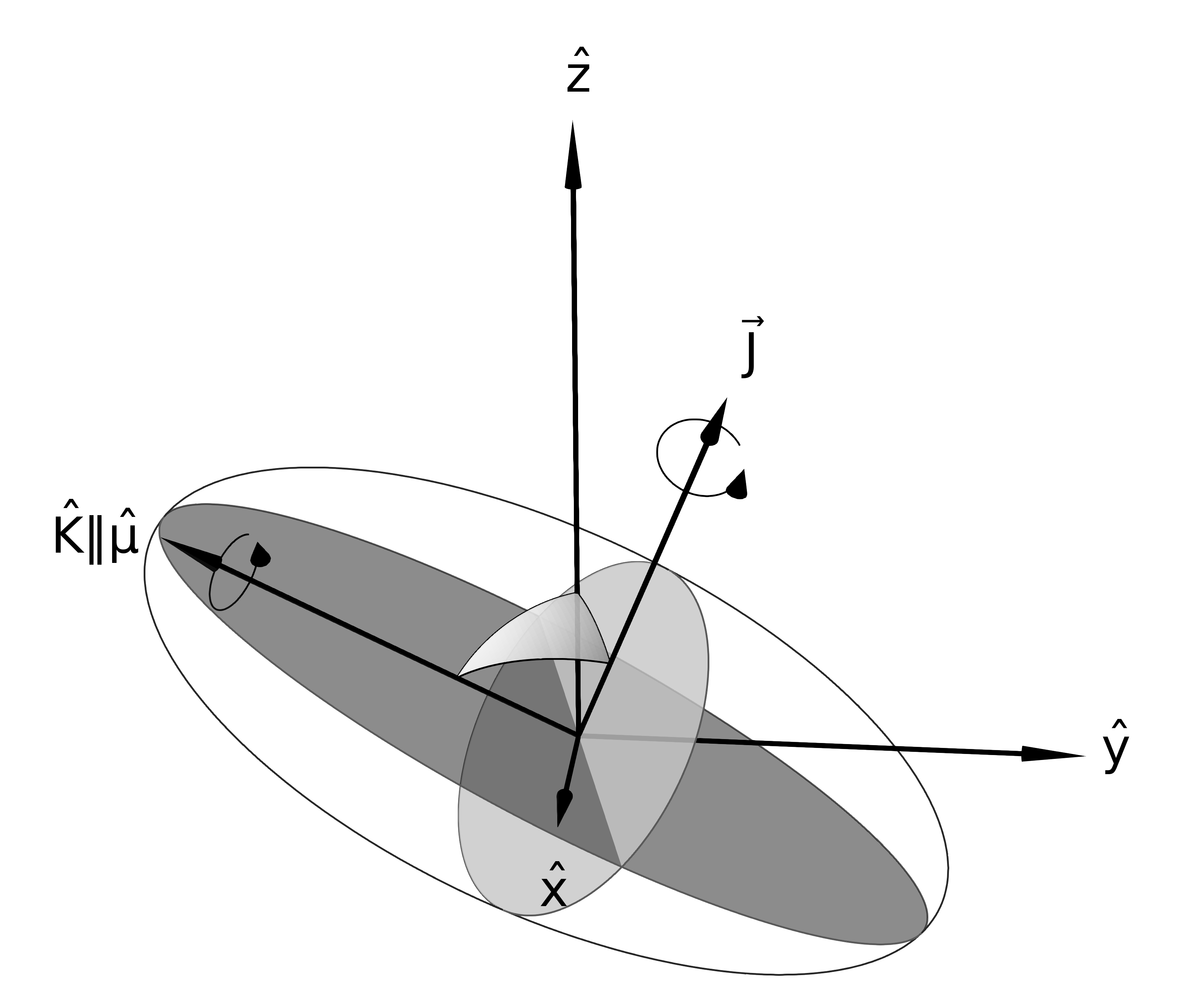}
    \caption{Relations between vectors relevant in rigid body rotations.
      $\hat{x}$-$\hat{y}$-$\hat{z}$ is the laboratory-fixed frame; $\vec{J}$ is the total angular momentum vector; $\hat{K}$ defines the figure axis, which coincides with the transition dipole $\hat{\mu}$ in our model.
      For clarity, we additionally plot the  $\hat{x}$-$\hat{K}$ axial section, the transverse section, and a spherical triangle spanned between $\hat{K}$, $\hat{z}$ and $\vec{J}$ vectors.
      \label{fig:rigid_rotor}}
  \end{figure}

  We start by describing $r_n(T)$ for a single H$_2$Pc-Ar$_n$ (or H$_2$Pc-Ne$_n$) cluster.
  We summarize the motion of linear rotors, symmetric tops and spherical tops~\cite{Herzberg1945,Taylor2005,Goldstein2008} to explain the difference between our results and those of Awali \textit{et al.}~\cite{Awali2021}.
  Figure~\ref{fig:rigid_rotor} illustrates the relations between key quantities in rigid body kinematics.
  The dipole autocorrelation function in Eq.~\eqref{eq:r-definition}, $\hat{\mu}(0)\cdot \hat{\mu}(T)$, is determined by torque-free dynamics of the cluster.
  For simplicity, we treat doped and undoped clusters as spherical tops.
  The effect of surface-doping is limited to equipping the cluster with a transition dipole $\vec{\mu}$ and increasing the moment of inertia, see Supp. Info. Sec.~\ref{app:hard-sphere}.
  For a spherical top, the moment of inertia around any axis going through the center of mass is equal, therefore figure axis $\hat{K}$ can be picked arbitrarily relative to the space-fixed total angular momentum axis $\vec{J}$.
  The motion of a cluster treated as a spherical top is particularly simple, because it rotates at a constant rate about $\vec{J}$.
  This corresponds to quantum energy level degeneracy with respect to $K$.
  On the other hand, in symmetric tops the choice is not arbitrary, degeneracy is lifted and the body additionally rotates around the figure axis.
  Note that the high symmetry of motion and increased degeneracy of spherical tops relative to symmetric tops does not reduce dimensionality of the problem.
  This is in contrast to linear rotors, in which the projection of total angular momentum onto the figure axis is fixed at zero ($K=0$), which limits $\hat{K}$ to a plane perpendicular to $\vec{J}$.
  This constraint is reflected in quantum and classical partition functions and in the frequency content of the anisotropy signal.

  We identify the figure axis with the transition dipole axis (of the dopant) and assume it remains fixed after excitation.
  In this case dynamics of the transition dipole in symmetric and spherical tops are the same, because the additional rotation of symmetric top around figure axis has no effect on $\hat{\mu}(0)\cdot \hat{\mu}(T)$ \footnote{Note that if we were to treat clusters as symmetric or asymmetric tops, then we would not be able identify the transition dipole axis with the figure axis, since the chromophore can attach to an arbitrary spot on the cluster.}.
  Within RCS theory these are known as parallel-parallel transitions and we can directly use previously derived formulas for anisotropy~\cite{Baskin1987}:
  \begin{equation}
    \label{eq:semi-symmetric}
    \begin{split}
       & r_{K,J}(T) = \frac{1}{10}\left(3\frac{K^2}{J^2} - 1\right)^2                                                             \\
       & + \frac{6}{10}\left(1 - \frac{K^2}{J^2}\right)\frac{K^2}{J^2} \left(\cos J\omega_{n0} T + \cos (J+1)\omega_{n0} T\right) \\
       & + \frac{3}{10}\left(1 - \frac{K^2}{J^2}\right)^2 \cos(2J+1)\omega_{n0}T,
    \end{split}
  \end{equation}
  where $\omega_{n0}=2B_n/\hbar$, $B_n$ is the rotational constant of Ar$_n$ cluster, which assumes zero-order expressions for rotational energy levels, \textit{e.g.} no centrifugal distortion in symmetric tops or Coriolis coupling in spherical tops.
  Note that the signal consists of a constant term and terms at $J\omega_{n0}$ and $2J\omega_{n0}$.
  Here we deviate from the model by Awali \textit{et al.}~\cite{Awali2021}, Eqs. (3) and (4) therein, which contain only constant and $2J\omega_{n0}$ terms. 
  We believe this difference arises from their use of Eq.~(20) of Ref.~\cite{Baskin1994}, which is appropriate for linear rotors but not for spherical tops.

  In the fully quantum case, the relative amplitude of terms in Eq.~\eqref{eq:semi-symmetric} depends not only on the orientation of the figure axis relative to the laboratory axis ($K/J$), but also on absolute $J$ and $K$ values.
  This dependence is only relevant for highly accurate, $J$-resolved amplitude measurements at low $J$ values, therefore it was omitted in Eq.~\eqref{eq:semi-symmetric}.
  We further approximate the signal by combining $J$ and $J+1$ terms and approximating $2J$ with $2J+1$, which yields:
  \begin{equation}
    \label{eq:semi-symmetric-2}
    \begin{split}
      r_{K,J}(T)  = & \frac{1}{10}\left(3\frac{K^2}{J^2} - 1\right)^2                                    \\
                    & + \frac{12}{10}\left(1 - \frac{K^2}{J^2}\right)\frac{K^2}{J^2} \cos J\omega_{n0} T \\
                    & + \frac{3}{10}\left(1 - \frac{K^2}{J^2}\right)^2 \cos 2J\omega_{n0}T.
    \end{split}
  \end{equation}
  We obtain the total anisotropy signal for a specific cluster size $n$ by averaging over the rotational state distribution:
  \begin{equation}
    \label{eq:semi-symmetric-total}
    r_n(T) = Q_n^{-1} \sum_{K,J} (2J+1) \exp \left[-E_{J,K}/k\Trot\right] r_{K,J}(T),
  \end{equation}
  where $Q_n$ is the rotational partition function, $k$ is the Boltzmann constant.
  For spherical tops in the zeroth order, rotational energies $E_{J,K}$ are independent of $K$, therefore by summing over $K$ the equation simplifies even further:
  \begin{widetext}
    \begin{equation}
      \label{eq:semi-spherical}
      r_n(T) = \frac{2}{25} + \frac{4}{25} Q_n^{-1} \sum_{J} 2J(2J+1) \exp \left[-B_n J(J+1)/k\Trot\right] ( \cos J\omega_{n0} T + \cos 2J\omega_{n0}T ).
    \end{equation}
  \end{widetext}
  The experimental signal is an average of Eq.~\eqref{eq:semi-spherical} over the size distribution of doped argon clusters $P(n)$:
  \begin{equation}
    \label{eq:semi-full}
    r(T) = \sum_n P(n) r_n(T).
  \end{equation}

  Cluster size distributions of various species have been found to follow the log-normal distribution~\cite{Wang1994,Lewerenz1993}.
  The log-normal distribution is consistent with the Hagena model, which assumes that clustering is dominated by growth due to sequential cluster-atom collisions and unimolecular decay due to single-atom evaporation~\cite{Hagena1987}.
  Moreover, the structural similarity between Ar$_n$ clusters of different sizes~\cite{Aziz1993,NAUMKIN1999,Bergersen2006} suggests an unimodal size distribution.
  The log-normal distribution was also used previously to model Ar$_n$ size distribution by Awali \textit{et al.}~\cite{Awali2021}.
  It is given by:
  \begin{equation}
    \label{eq:log-normal}
    P_{\mathrm{Ar}}(n) = \frac{1}{n} \frac{1}{\sigma\sqrt{2\pi}}
    \exp\left( -\frac{(\ln n - \mu)^2}{2\sigma^2} \right).
  \end{equation}
  The distribution is characterized by parameters $\mu$ and $\sigma$, which are non-trivially related to the mean and variance of the cluster size:
  \begin{subequations}
    \label{eq:log-normal-params}
    \begin{eqnarray}
      \langle n \rangle & = & e^{\mu+\sigma^{2}/2},\label{eq:log-normal-mean}\\
      \Var(n) & = & (e^{\sigma^2}-1)e^{2\mu+\sigma^2}.\label{eq:log-normal-var}
    \end{eqnarray}
  \end{subequations}
  The size distribution of doped clusters is shifted towards larger $n$ by the pick-up of H$_2$Pc molecules, since larger clusters are more likely to encounter the dopant.
  Assuming a simple hard-sphere collision kernel, $P_{\mathrm{Ar}}(n)$ needs to be scaled by the geometrical factor $n^{2/3}$, from which we obtain:
  \begin{equation}
    \label{eq:log-normal-doped}
    P(n) = \exp\left[ -\frac{2}{3}\left(\mu+\frac{1}{3}\sigma^2\right) \right]
    n^{2/3} P_{\mathrm{Ar}}(n),
  \end{equation}
  as the distribution of doped clusters.
  This allows us to evaluate Eq.~\eqref{eq:semi-full}.

  A naive calculation of the experimental signal involves a nested sum over all experimental delay times, all cluster sizes with appreciable population, and all rotational states with non-negligible occupation.
  Despite simplicity of the model, this calculation can be time-consuming.
  The computation time can be reduced by orders of magnitude if one notices that Eq.~\eqref{eq:semi-spherical} is a sum of two inverse discrete cosine transforms from the frequency domain in units of $J\omega_{n0}$ and $2J\omega_{n0}$ to the time domain.
  We exploit this fact by first calculating the full signal spectrum for the whole cluster distribution and only then perform an inverse transform with an optimized FFT code~\cite{Frigo2005,Harris2020,Virtanen2020}.

  We further speed up the calculations by exploiting another feature of the signal.
  The comb-like structure of rotational anisotropy spectrum, Eq.~\eqref{eq:semi-spherical}, with $J$ numbering individual comb lines, implies a pulse train in the time domain.
  After initial dephasing, the transition dipoles rephase every $T_r=1/(4B)$.
  For Ar$_n$ clusters, this period scales with size as:
  \begin{equation}
    \label{eq:rephasing-period}
    T_r^{(n)} \approx 0.0341\text{ ns} \times n^{5/3},
  \end{equation}
  hence for $n>=13$ revivals occur beyond our maximum experimental signal length of $T_{\mathrm{max}}=2.5$ ns.
  Moreover, since the signal is produced by a broad distribution of cluster sizes, it spans many octaves of the revival time constant, therefore even for longer observation times rotational revivals would largely average out.
  As a consequence, the quantum nature of the signal is lost and the discrete sum over $J$ numbers can be replaced by an integral without any loss of fidelity.
  Therefore, it is sufficient to sample the semi-classical signal spectrum, see Sec. \ref{app:fdomain}, in frequency steps of $1/T_{\mathrm{max}}$ up to $1/(2\Delta T)$, where $\Delta T$ is the experimental step size.
  With both improvements, we reduce the computation time by $10^5$ compared to a naive sum.

  \section{\label{sec:Results}Results}
  \begin{figure*}
    \centering
    \includegraphics[width = 0.7\linewidth]{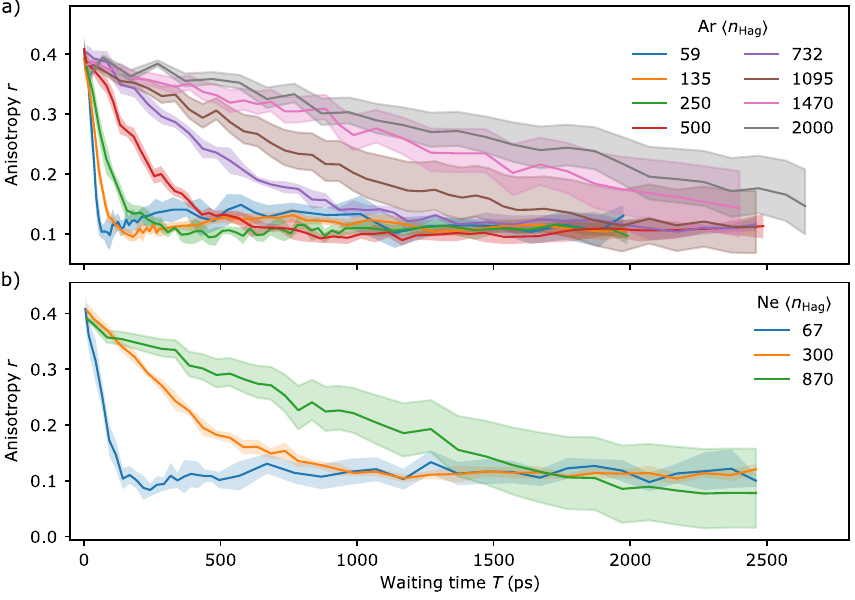}
    \caption{\label{fig:exp_data} Experimental data of the temporal anisotropy dephasing for different argon (a) and neon (b) clusters sizes. $\langle n_\mathrm{Hag} \rangle$ labels indicate the mean cluster size predicted by the Hagena scaling laws. Shaded areas indicate the standard deviation of the measurement. The data are scaled to match the theoretically expected asymptotic behavior \cite{Baskin1987}.}
  \end{figure*}
  Fig.~\ref{fig:exp_data}a,b show the anisotropy parameter $r(T)$ reconstructed from the experimental data for different Ar and Ne mean cluster sizes. 
  Clearly, the anisotropy dephasing time decreases with decreasing mean cluster size.
  Remarkably, even small differences of a few tens of atoms in mean cluster size can be distinguished.
  The observed trend can be rationalized by the fact that smaller clusters rotate faster leading to faster rotational decoherence.
  Likewise, the faster anisotropy dephasing for Ar compared to Ne clusters is in accordance with the lower temperature of Ne compared to Ar clusters \cite{Farges1980}.

  \begin{figure*}
    \centering
    \includegraphics[width=0.7\linewidth]{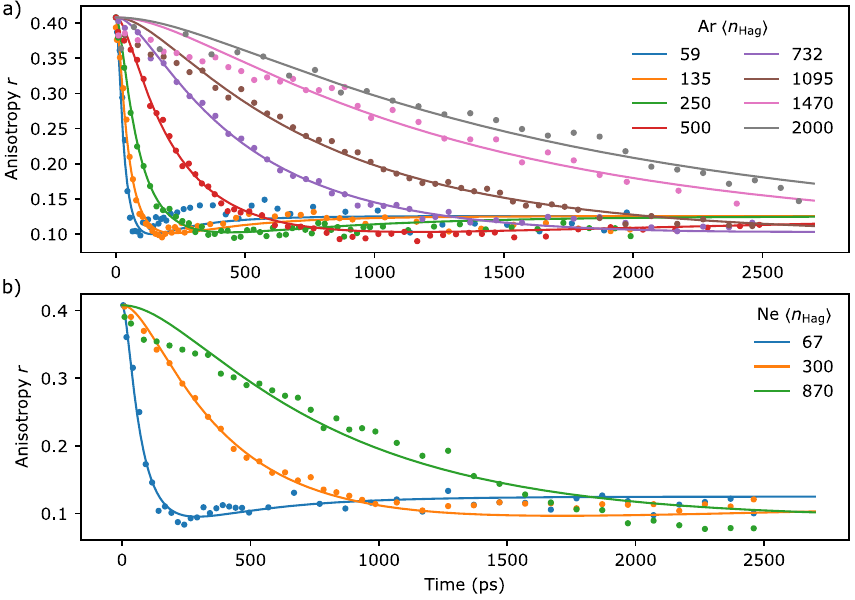}
    \caption{\textbf{a} Experimental data that was shown in Fig.~\ref{fig:exp_data}a (dots) compared to the model presented in Sec.~\ref{sec:theory} (solid lines) for Ar clusters. Experimental uncertainties are not shown for better visibility. \textbf{b} Same for Ne cluster data shown in Fig.~\ref{fig:exp_data}b. Labels indicate the cluster sizes predicted by the Hagena scaling laws.}
    \label{fig:Fig_comp_data_fits}
  \end{figure*}
  In Fig.~\ref{fig:Fig_comp_data_fits} the data are compared to the model discussed in Sec.~\ref{sec:theory}.
  The model is scaled to the contrast between maximum and minimum of the experimental data and a least-square fit is performed using the mean cluster size and width of the size distribution as fit parameters.
  The initial scaling step is necessary to account for residual ellipticity in the laser polarization \cite{Baskin1987} (see SI Fig.~\ref{fig:comp_scaling_SI} for details) and does not influence the anisotropy dephasing time which determines the mass of the clusters.

  The comparison between experiment and theory shows excellent agreement, confirming the validity of the theoretical model and the ability to deduce mean cluster sizes from the data.
  An exception applies for the smallest Ar clusters corresponding to a Hagena size estimate of $\langle n_{\mathrm{Hag}} \rangle = 59$. Here, the data and model exhibit a mismatch at early decay times below 500\,ps, which is attributed to a mismatch in the rotational temperature between experiment and theory.
  For small cluster masses $\langle n \rangle < 100$, the doping with the relatively heavy molecules may transfer significant angular momentum during the doping process, which effectively increases the rotational temperature of the clusters \cite{Awali2021}.
  This effect is neglected in our theoretical model for simplicity.
  Conversely, for larger clusters, angular momentum transfer can be well neglected due to the large cluster/dopant mass ratio.
  Note, that this issue is not observed for the Ne clusters, which is due to the fact, that these clusters are much larger than predicted by the Hagena scaling laws (see below).

  \begin{figure}
    \centering
    \includegraphics[width = \linewidth]{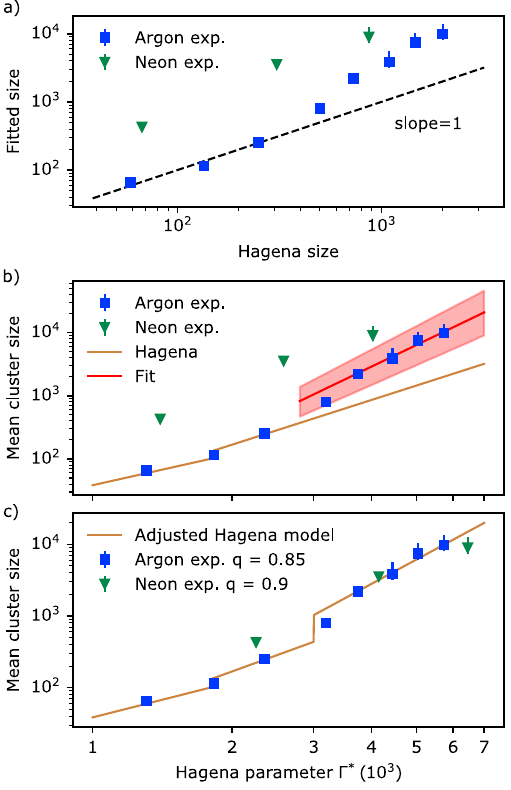}
    \caption{\label{fig:resulting_cluster_sizes}\textbf{a} Cluster size determined by rotational coherence spectroscopy (y-axis) compared to the Hagena scaling law (x-axis). Blue squares and green triangles show the experimentally determined cluster sizes. \textbf{b} Cluster size as a function of the Hagena parameter $\Gamma^*$: Hagena scaling law (beige line), our study (blue squares, green triangles). The Ar cluster data for $\Gamma^* > 3000$ is fitted with the model in Eq.\,\ref{eq:adjust} (red), empirical confidence interval (red shaded area). \textbf{c} Same as \textbf{b} but with an adjusted Hagena model as discussed in the main text.}
  \end{figure}
  The cluster sizes obtained from the comparison between the data and the model are shown in Fig.~\ref{fig:resulting_cluster_sizes}.
  These results have been obtained for $\sigma_\mathrm{Ar} = 1.0$ and $\sigma_\mathrm{Ne}  =0.8$ as widths for the cluster size distribution of the Ar and Ne cluster beams, which agree well with previously recorded broad cluster size distributions \cite{Schutte2002}. In the case of Ar, it can be clearly seen in Fig.~\ref{fig:resulting_cluster_sizes}a that our results agree with the Hagena scaling laws for small mean cluster sizes.
  However, with increasing cluster size, the Hagena scaling laws underestimates the mean cluster size by a factor of two to four, which matches with other studies \cite{Farges1986,Wormer1989,Lee2020}. In the case of Ne, an even greater deviation by almost an order of magnitude is observed (Fig.~\ref{fig:resulting_cluster_sizes}a). This will be discussed further below and the following discussion will focus on the Ar clusters.

  Fig.~\ref{fig:resulting_cluster_sizes}b shows our data as a function of the Hagena parameter $\Gamma^*$ along with the Hagena scaling law based on Eq.~\ref{eq:Hagena} and Eq.~\ref{eq:Buck}.
  Excellent agreement between our study and the Hagena scaling law is obtained for the first three data points, while for $\Gamma^* > 3000$ the Hagena model starts to underestimate our measurements as previously observed. In order to explain these data better, an equation in the form of the Hagena scaling law
  \begin{align}
    \label{eq:adjust}
    \langle n \rangle = a \times \left(\dfrac{\Gamma^*}{1000}\right)^b
  \end{align}
  is fitted to the data with $\Gamma^* > 3000$. The resulting curve with $a = 22 \pm 5$ and $b = 3.5 \pm 0.3$ is shown in Fig.~\ref{fig:Fig_comp_data_fits}b and agrees well with the data. We propose this as a new scaling regime to predict the cluster size of larger clusters more accurately. The combined scaling curve based on Eq.~\ref{eq:Hagena}, Eq.~\ref{eq:Buck} and Eq.\,\ref{eq:adjust} using $a = 22, b = 3.5$ is shown in Fig.~\ref{fig:resulting_cluster_sizes}c.
  We note, that we used here for the empirical parameter $q$ of the Hagena scaling law (see Eq.\,\ref{eq:gamma}) a value of 0.85, in accordance with the suggestion in Ref.\cite{Hagena1987} for Ar clusters.

  While this procedure works well for the Ar clusters, we observe still a significant systematic discrepancy for Ne clusters.
  We therefore propose an adjustment of $q$.
  The theoretical limits for $q$ are $0.5 < q \le 1$ \cite{Hagena1987}.
  Experiments with Ar cluster beams yielded values in the range from $q = 0.8$ \cite{Hagena1972,Hagena1974} up to $q = 0.93$ \cite{Habetsphd}, based on which $q= 0.85$ for Ar clusters was suggested~\cite{Hagena1987}.
  It is sensible that other rare gases such as Ne should have a similar value for $q$, albeit not necessarily the same. Using a value of $q_{\mathrm{Ne}} = 0.9$ for Ne, for example, gives a much better agreement between the measured Ne mean cluster size and the adjusted scaling, as shown in Fig.~\ref{fig:resulting_cluster_sizes}c.

  The strong sensitivity to the exact value of $q$ is most likely a result of different expansion conditions between the Ar and Ne experiments. In Ref.~\cite{Hagena1987} it is argued that the exact value of $q$ should not affect the comparison between different gases as long as the effective nozzle opening and the expansion temperature are on the same order of magnitude. In the experiments conducted in this work, the product of nozzle diameter and expansion temperature $d\cdot T_{\mathrm{cl}}$ is larger by a factor of six to ten for the Ar experiments compared to Ne (Tab.\,\ref{tab:cluster_para}). Thus, it is plausible that the exact value of $q$ affects our results significantly.

  \section{Discussion}
  While characterizing the temporal polarization anisotropy dephasing of an ensemble of doped clusters omits ionization and thus possible fragmentation of the clusters, other factors may lead to systematic errors in the cluster size determination using this approach.

  Our theoretical model assumes full thermalization between the rotational and vibrational degrees of freedom of the clusters and hence equal rotational and vibrational cluster temperatures.
  The vibrational cluster temperatures were previously determined as $T_\mathrm{Ar} = 37$\,K and $T_\mathrm{Ne} = 10$\,K for Ar and Ne clusters, respectively \cite{Farges1980}.
  The vibrational temperature is efficiently reduced by the binding energy of each evaporated rare-gas atom, whereas the change in angular momentum on average is small due to the isotropic evaporation of rare-gas atoms.
  Hence, it is likely that during formation of the rare-gas clusters and subsequent evaporative cooling their rotational temperature relaxes more slowly towards equilibrium than the vibrational temperature\cite{Pauly2000}.
  Likewise, our model neglects the possible angular momentum transfer during the doping process (see discussion above).
  While this is well justified for large cluster/dopant mass ratios, we observed for smaller clusters a mismatch between the data and model which we attribute to this effect (Fig.\,\ref{fig:Fig_comp_data_fits}a).

  Both effects, the lacking equilibration between rotational and vibrational temperatures and the neglection of angular momentum transfer in the doping of small clusters, can result in an underestimation of the rotational cluster temperature underlying our theoretical model.
  This would result in a systematic error in the determined cluster sizes.
  The anisotropy dephasing time scales $\propto \sqrt{m/\Trot}$, where $m$ denotes the mass of the rotor\,\cite{Felker1987}.
  Hence, an underestimate of the rotational temperature results in an underestimate of the mean cluster size. 
  For these reasons, we conclude that our results may exhibit a small systematic uncertainty and the real cluster sizes might be higher than the size determined in our work.

  We note, that the evaporation of rare-gas atoms from the clusters due to doping with the hot H$_2$Pc molecules leads to a shrinking of the initial cluster size distribution.
  In principle, this might add another systematic error source for the cluster size estimation.
  However, we calculated that about 40 Ne and 10 Ar atoms are evaporated, respectively, upon doping of one H$_2$Pc molecule.
  This seems negligible for most cluster sizes investigated in this work, which is why we have not included this contribution in our model.
  However, for the smallest investigated Ar cluster size, this effect may also contribute to the discrepancy between the experiment and the model.


  \section{\label{sec:Conclusion}Conclusion}
  We presented a systematic study of an ionization-free method to determine the size of rare-gas clusters based on measuring the temporal polarization anisotropy dephasing of an ensemble of surface-doped clusters.
  We developed an efficient analytical model to describe the polarization anisotropy signal with which we extracted the experimental mean cluster sizes.
  Our approach shows a remarkable sensitivity to small changes in the mean cluster size and the good agreement between experiment and theory validates our approach.
  In particular, for larger clusters containing more than $\approx 500$ atoms, we find a significant systematic deviation from the Hagena scaling law and provide a correction based on our experimental data.
  We also show the importance of the empirical $q$-parameter in the Hagena scaling law and provide adequate numbers for $q$ for Ar and Ne clusters.

  Our proposed correction of the Hagena scaling law is general and may be used for more accurate size estimates of rare-gas clusters produced in continuous-wave jet expansion experiments.
  The demonstrated approach may be also used to further refine the Hagena scaling law for even larger clusters than the ones studied here.
  Likewise, clusters of other rare-gas species may be analyzed using our approach to gain more accurate estimates of the $q$-parameter.
  Further experiments would be needed for size estimates of clusters produced in pulsed jet expansion.
  The same applies for the size characterization of superfluid helium nanodroplets constituting a special class of rare-gas clusters.
  However, due to the superfluid nature of these clusters, the here presented approach cannot be applied and other methods are needed.

  \section{Funding}
  Funding from Deutsche Forschungsgemeinschaft RTG 2717. This work is also based upon the work of COST Action CA21101 "Confined molecular systems: from a new generation of materials to the stars" (COSY) supported by COST (European Cooperation in Science and Technology).

  \section{Acknowledgment}
  We acknowledge fruitful discussions with Bernd von Issendorff.

  \section{Disclosures}
  The authors declare no conflicts of interest.

  \section{Data Availability Statement}
  The experimental and simulation data included in this work are available on the open repository: \textit{Accession codes will be available before publication}.

  \section{\label{sec:SI}Supporting information}


  \subsection{\label{app:hard-sphere}Cluster moments of inertia}
  Following Awali \textit{et al.}~\cite{Awali2021}, we model Ar$_n$ and He$_n$ as homogeneous spheres.
  The moment of inertia is given by $I_n^{(\mathrm{at})}=\frac{2}{5}nM_{\mathrm{at}}R_n^2$, where $M_{\mathrm{at}}$ is the atomic weight of the element. $R_n=R_{\mathrm{at}} \left(\frac{n}{c}\right)^{1/3}$ is the cluster radius, where $c=0.74$ is the atomic packing factor for face-centered cubic lattice and  $R_{\mathrm{at}}$ is the hard-sphere radius of the element.
  In this work, we used:
  \begin{itemize}
    \item $M_{\mathrm{Ar}} = 39.95$, $R_{\mathrm{Ar}} = \SI{188}{\pico\meter}$,
    \item $M_{\mathrm{Ne}} = 20.1797$, $R_{\mathrm{Ne}} = \SI{154}{\pico\meter}$,
    \item $M_{\mathrm{H_2Pc}} = 514.552$.
  \end{itemize}

  The mass of H$_2$Pc is assumed to be uniformly distributed over the cluster surface, such that the moment of inertia of a doped cluster is given by:
  \begin{equation}
    \label{eq:doped-inertia}
    I_n = (\frac{2}{5}nM_{\mathrm{at}} + \frac{2}{3}M_{\mathrm{H_2Pc}})R_n^2.
  \end{equation}

  \subsection{\label{app:fdomain}Frequency-domain anisotropy signal in the classical limit}
  Fast calculation of anisotropy signals is facilitated by changing sums over quantum numbers into integrals that can be sampled at arbitrary points in $f$-domain.
  We rewrite Eq.~\eqref{eq:semi-spherical} as:
  \begin{equation}
    \label{eq:eq:semi-spherical-cont}
    r_n(T) = \frac{2}{25} + \int_{-\infty}^{\infty}\mathrm{d}f [A_1(f) + A_2(f)]e^{2\pi f T},
  \end{equation}
  \textit{i.e.} an inverse Fourier transform between \textit{T}-\textit{f} Fourier pair.
$A_{1}(f)$ and $A_{2}(f)$ correspond to $J\omega_{n0}$ and $2J\omega_{n0}$ terms and can be expressed as:
  \begin{subequations}
    \label{eq:A1-A2}
    \begin{eqnarray}
      A_1(f) & = & 16C_1 (\Trot B)^{-3/2} f^2 \exp \left[ -\frac{f^2}{C_2B\Trot} \right],\label{eq:A1}\\
      A_2(f) & = & C_1 (\Trot B)^{-3/2} f^2 \exp \left[ -\frac{f^2}{4C_2B\Trot} \right],\label{eq:A2}
    \end{eqnarray}
  \end{subequations}
  where $C_1 = \frac{1}{200}\frac{h^3}{\sqrt{\pi}k^{3/2}}$, $C_2 = 4k/h^2$.

  \subsection{\label{app:derivation_ani}Derivation of corrections to the rotational-coherence-spectroscopy signal}
  When recording an anisotropy signal from fluorescence, the excitation and detection geometry need to be accounted for. In the following, the derivation of the corrections are presented that need to be employed for the geometry in this work (see Fig.~\ref{fig:experimental_scheme}c). They are based on the work of Refs.~\cite{Baskin1994,Brown1999}, where more details can be found.

  In the following, it will be assumed that the doped rare-gas clusters can be described as spherical symmetric tops. Based on this, a signal $S_{\mathrm{sig}, i}$ that is emitted by the excited sample in the direction $i$ $(i \in \left\{ x,y,z \right \})$ is given by
  \begin{widetext}
    \begin{align}
      \label{eq:signal_dir}
      S_{sig, i}(\tau) = \langle \langle \langle \left( \bm{\epsilon}_1 \bm{d}(\tau = 0) \right)^2 \left( \bm{\epsilon}_2 \bm{d}(\tau) \right)^2 \langle \left( \bm{\epsilon}_{sig,i} \bm{d}(\tau + t^\prime) \right)^2 \rangle_{t^\prime} \rangle_{\psi_0}\rangle_{(\theta, \phi)_{4\pi}}\rangle_{(\theta_p, \phi_p)_{4\pi}}.
    \end{align}
  \end{widetext}
  Here, the excitation pulse is polarized along $\bm{\epsilon}_1$, the probe pulse is polarized along $\bm{\epsilon}_2$ and afterwards light is emitted in the direction $\bm{\epsilon}_{sig, i}$ ($i = \{x,y,z\}$). The interaction of the pump, the probe pulse, and the emission direction of the excited molecule are averaged over the lifetime of the emitter $t^\prime$, the initial orientation of the molecules $\psi_0$, over the solid angles ($\theta, \phi$) and over the rotation angles ($\phi_p, \theta_p$) between the frame of reference for polarization and angular momentum. If the lifetime is long compared to the rotational period, which is the case in the experiment, averaging over $t^\prime$ can be appodized to only one rotational period. The initial rotation $\psi_0$ is averaged over $2 \pi$ and ($\theta, \phi$) as well as ($\phi_p, \theta p$) are averaged over $4\pi$.

  The excitation pulses interact with the transition dipole moment $\bm{d}(t)$ of the dopant molecules, which is given by
  \begin{align}
    \bm{\hat{e}}_x \bm{d}(t) & = \sin \phi_p \sin \theta \cos \theta_p \cos\psi(t)                          \\
                             & +\sin\phi_p\sin\theta_p \cos\theta + \sin\theta \sin\psi(t)\cos\phi_p \notag \\
    \bm{\hat{e}}_y \bm{d}(t) & = \sin\phi_p\sin\theta\sin\psi(t)                                            \\                             & -\sin\theta\cos\phi_p\cos\theta_p\cos\psi(t) - \sin\theta_p\cos\phi_p\cos\theta \notag \\
    \bm{\hat{e}}_z \bm{d}(t) & = -\sin\theta\sin\theta_p\cos\psi(t) + \cos\theta\cos\theta_p.
    \label{eq:TDM_orient}
  \end{align}
  Here, $\psi(t)$ describes the time evolution of the orientation of the dipole moment, which is given by
  \begin{equation}
    \psi(t) = \omega t + \psi_0,
  \end{equation}
  for a constant angular frequency $\omega$ and an initial orientation $\psi_0$ at time $t = 0$. Eq.~\ref{eq:TDM_orient} is different than what is typically derived in literature, for example, in Ref.~\cite{Baskin1994}. This is due to the fact that rotations of spherical symmetric tops are considered in this work, where literature typically deals with linear rotors.

  With all of this established, Eq.~\ref{eq:signal_dir} can be solved and yields in the case of excitation pulses with the same polarization, i.e. signal $I_\parallel$ with $\bm{\epsilon}_1 = (0,0,1)$ and $\bm{\epsilon}_2 = (0,0,1)$:
  \begin{align}
    S_x^{(\parallel)}(\tau) = S_y^{(\parallel)}(\tau) & = \frac{403 + 129\cos(\omega \tau) + 120 \cos(2\omega\tau)}{11025} \\
    S_z^{(\parallel)}(\tau)                           & = \frac{615+136\cos(\omega\tau) + 152\cos(2\omega\tau)}{11025}.
  \end{align}
  For the geometry of $I_\perp$, where $\bm{\epsilon}_1 = \frac{1}{\sqrt{2}}(1,0,1)$ and $\bm{\epsilon}_2 = \frac{1}{\sqrt{2}}(1,0,-1)$, Eq.~\ref{eq:signal_dir} yields
  \begin{align}
    S_x^{(\perp)}(\tau) = S_z^{(\perp)}(\tau) & = \frac{403-68\cos(\omega\tau) - 76 \cos(2\omega\tau)}{11025} \\
    S_y^{(\perp)}(\tau)                       & = \frac{321-60\cos(\omega\tau) - 44\cos(2\omega\tau)}{11025}.
  \end{align}
  Due to the limited solid angle mapped by the detector, only a portion of the above-calculated signals is acquired. For example, in case of a point-like detector placed in $\bm{\hat{e}}_x$ direction from the sample, the resulting signal would be
  \begin{equation}
    I^{(\parallel/\perp)} = S_y^{(\parallel/\perp)}(\tau) + S_z^{(\parallel/\perp)}(\tau).
  \end{equation}
  This is, however, not the case for the experiments presented in this study, where the fluorescence is mapped on to a photomultiplier with a numerical aperture of $N_A = 0.76$. This translates into an opening angle $\theta^\prime = 49.46$ $^\circ$ using $\sin \theta^\prime = N_A$. Based on this opening angle, the portion of the light emitted in the x-direction can be calculated from the overlap between the opening angle of the detector with the radiation $P(\theta)$ from a dipole in the direction $\theta$, which is given by $P \propto \sin^2(\theta)$. Thus, the fraction of the signal in x-direction in dependence of the opening angle $\theta$ is obtained from
  \begin{align}
    F(\theta) & = \int_0^{\theta} d\theta^\prime \sin^3\theta^\prime \notag     \\
              & = \frac{4}{3}(2+\cos\theta)\sin^4\left(\frac{\theta}{2}\right).
  \end{align}
  With $\theta^\prime = 49.46$, this yields that $\approx$10\,\% of the fluorescence emitted in x-direction is collected by the employed detection geometry.

  With this result, the total signal collected by the detector can be calculated from
  \begin{align}
    \label{eq:int_total}
    I^{(\parallel/\perp)} = 0.1\cdot S_x^{(\parallel/\perp)}(\tau) + S_y^{(\parallel/\perp)}(\tau) + S_z^{(\parallel/\perp)}(\tau),
  \end{align}
  which, expressed with $S_i^{(\parallel/\perp)}$, is given by
  \begin{align}
    \label{eq:int_values}
    I_\parallel & = \frac{7743+2768\cos(\omega\tau) + 5680\cos^2(\omega\tau)}{110250} \\
    \mathrm{and} \notag                                                               \\
    \label{eq:int_values2}
    I_\perp     & =\frac{8919-1348\cos(\omega\tau) -2552\cos^2(\omega\tau)}{110250}.
  \end{align}
  Here, the $\cos(2\omega\tau)$ terms were transformed into $\cos^2(\omega\tau)$ with the relation $2\cos^2(\omega\tau) = 1+\cos(2\omega\tau)$. Based on this, the anisotropy signal $r(t)$ can be calculated from
  \begin{align}
    r(t) = \dfrac{I_{\parallel} - I_\perp}{I_{iso}},
  \end{align}
  where $I_{iso}$ denotes the total isotropic signal to which the signal is normalized. Typically, it is given by
  \begin{align}
    \label{eq:Iso_org}
    I_{iso} = I_\parallel + 2 I_\perp.
  \end{align}
  However, due to the imbalanced detection of the different polarization contributions, Eq.~\ref{eq:Iso_org} does not yield the correct isotropic contribution for normalization. The authors in Ref.~\cite{Brown1999} argue that this can be easily seen through the fact that plugging the intensities from Eqs.~\ref{eq:int_values},\ref{eq:int_values2} into Eq.~\ref{eq:Iso_org} does not lead to a cancellation of all $\cos^2$-terms with ``rotational'' content. Only if these terms cancel out, the ``pure'' isotropic contribution is obtained. Thus, according to Ref.~\cite{Brown1999} the correct formulation of Eq.~\ref{eq:Iso_org} can be simply found by
  \begin{equation}
    I_{iso} = I_\parallel + A I_\perp,
  \end{equation}
  with an appropriate choice of the factor $A$. For the geometry of this work, $A = 2.23$ satisfies the above-discussed conditions.

  \begin{figure*}
    \centering
    \includegraphics[width=0.7\linewidth]{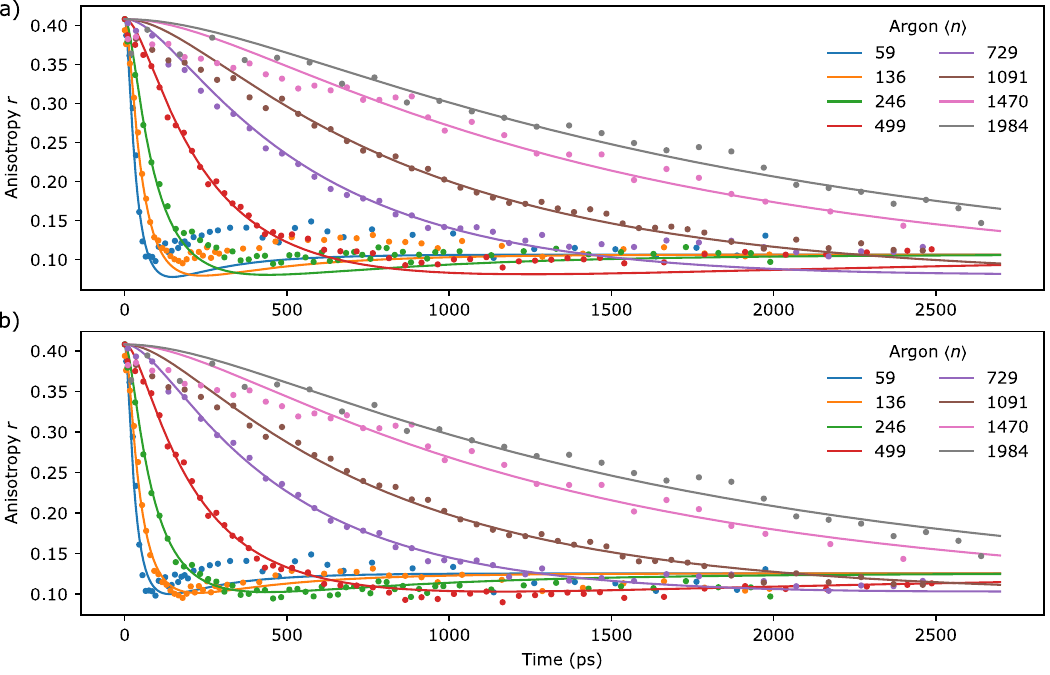}
    \caption{\textbf{a} Experimental anisotropy signal of H$_2$Pc on argon clusters (dots) compared to fits (solid lines) based on the model presented in the main text in Sec.~\ref{sec:Methods}. \textbf{b} Same as \textbf{a} but with the model's contrast scaled to the experimental contrast to account for residual ellipticity in the beam.}
    \label{fig:comp_scaling_SI}
  \end{figure*}
  It should be noted, that this treatment neglects the contribution of the $\cos$-terms. This is done because they do not contribute for $t \rightarrow \infty$ and their contribution is close to zero for $A = 2.23$ (exact cancellation would be achieved with $A_2 = 2.05$).

  Considering all these aspects, it can be concluded that
  \begin{align}
    \label{eq:ani_corrected_app}
    r(t) = \dfrac{I_\parallel - I_\perp}{I_\parallel + 2.23 \cdot I_\perp}
  \end{align}
  gives the correct anisotropy value for the signals obtained with the excitation and detection geometry employed in this work. An interesting implication of Eqs.~\ref{eq:int_values},\ref{eq:int_values2},\ref{eq:ani_corrected_app} is that the anisotropy value at $t = 0$ and $t = \infty$ differ from what is typically given in the literature. With Eq.~\ref{eq:Iso_org}, values of $r_\mathrm{lit}(t = 0) = 0.4$ and $r_\mathrm{lit}(t = \infty) = 0.1$ are obtained \cite{Baskin1994,Lakowicz2006,Valeur2012}. For the geometry of this work, however, values of
\begin{align}
  r(0)      & = 0.41  \\
  r(\infty) & = 0.11,
\end{align}
are obtained. This changes the scaling of the data, which is shown in Fig.~\ref{fig:comp_scaling_SI}. Here, the contrast of the model discussed in this study was scaled to the anisotropy contrast of the experiment to account for residual ellipticity in the experiment similar to Ref.~\cite{Baskin1987}.

\clearpage
\bibliography{Sources_anisotropy}

\end{document}